\documentclass[a4paper,11pt]{article}
\usepackage{pos}
\usepackage{color} 

\definecolor{green}{rgb}{0,.5,0}

\definecolor{red}{rgb}{1,0,0}

\usepackage{graphicx}
\usepackage[subfigure]{graphfig}
\usepackage{epsfig}
\usepackage{slashed}
\usepackage{dcolumn}
\usepackage{amsmath}
\usepackage{latexsym}
\usepackage{multirow}

\newcommand{\GeV}{\text{GeV}}

\newcommand{\MOM}{\text{RI/MOM}}

\newcommand{\SMOM}{\text{RI/SMOM}}
\newcommand{\MSbar}{{\overline{\text{MS}}}}

\def\be{\begin{equation}}
\def\ee{\end{equation}}
\def\bea{\begin{eqnarray}}
\def\eea{\end{eqnarray}}
\def\non{\nonumber}

\title{Renormalization constants of overlap quark bilinear operators from \MOM\ and \SMOM~scheme}

\author[a]{Fangcheng He}
\author*[a,b,c,d]{Yi-Bo Yang}

\affiliation[a]{CAS Key Laboratory of Theoretical Physics, Institute of Theoretical Physics,\\
  Chinese Academy of Sciences, Beijing 100190, China}

\affiliation[b]{School of Fundamental Physics and Mathematical Sciences, Hangzhou Institute for Advanced Study,\\
UCAS, Hangzhou 310024, China}

\affiliation[c]{International Centre for Theoretical Physics Asia-Pacific, Beijing/Hangzhou, China}

\affiliation[d]{School of Physical Sciences, University of Chinese Academy of Sciences, Beijing 100049, China
}

\emailAdd{hefc123@itp.ac.cn}
\emailAdd{ybyang@itp.ac.cn}

\abstract{We calculate the renormalization constants (RCs) of vector, axial, vector scalar, pseudoscalar and tensor quark operators of the overlap valence fermion, on the 11 gauge ensembles with dynamical fermion using either Domain wall fermion~(DWF) action or
Highly improved stagger quark~(HISQ) action at lattice spacings from 0.04 fm to 0.20 fm. We find the results under the $\MSbar$ scheme using either the $\MOM$ or $\SMOM$ scheme are consistent with each other, once the  proper $a^2p^2$ extrapolation is applied and the systematic uncertainties are estimated cautiously. 
Our results with different gauge and fermion actions also indicate that the RCs
are majorly dependent on the lattice spacing (as the inverse of UV cut-off) rather than the bare gauge
coupling used by the gauge action.}

\FullConference{%
 The 38th International Symposium on Lattice Field Theory, LATTICE2021
  26th-30th July, 2021
  Zoom/Gather@Massachusetts Institute of Technology
}

\begin{document}
\maketitle

\section{Introduction}
$\MOM$~\cite{Martinelli:1994ty} and $\SMOM$~\cite{Aoki:2007xm,Sturm:2009kb} schemes are widely used to renormalize the bare matrix element calculated by Lattice QCD.
In the $\MOM$ scheme, the vertex correction is considered in the forward off-shell parton state.
The momenta of external quark legs are 
chosen to be $p_1=p_2=p$ with the renormalization scale $\mu$ defined by $\mu^2=p^2$, and the RCs are determined by the following renormalization conditions~\cite{Martinelli:1994ty},
\begin{subequations}
\label{eq:rimom}
\begin{eqnarray}
\label{eq:rimom_prop}
Z_q^{\MOM}&=&\mathop{\textrm{lim}}\limits_{m_R\rightarrow0}\frac{-i}{48}\textrm{Tr}\Big[\gamma_\mu\frac{\partial S_B^{-1}(p)}{\partial p_\mu}\Big]_{p^2=\mu^2}, 
\label{eq:rimom_Zo}\non\\
Z^\MOM_{\mathcal{O}}&=&
\mathop{\text{lim}}\limits_{m_R\rightarrow0}\frac{Z_q^\MOM}{\frac{1}{12}\text{Tr}[\Lambda_{\mathcal{O},B}(p,p)\Lambda^{\text{tree}}_{\mathcal{O}}(p,p)^{-1}]}\Big |_{p^2=\mu^2}.
\end{eqnarray}
\end{subequations}
However, the definition in Eq.~(\ref{eq:rimom_prop}) introduces the derivative to the momentum, which is cumbersome to fulfill in the discretized lattice. 
A better choice to obtain $Z_q^{\MOM}$ is through the vector vertex correction
\begin{equation}\label{eq:rimom_vex_zq}
Z_q^{\MOM}=\mathop{\text{lim}}\limits_{m_R\rightarrow0}\frac{Z_V^\MOM}{12}\text{Tr}[\Lambda^\mu_{V,B}(p,p)\gamma_\mu],
\end{equation}
where $\Lambda_{V,B}(p,p)$ is the bare vector vertex.  Thus the RCs of arbitrary quark bi-linear operators $\mathcal{O}$ 
can be obtained through
\begin{equation}\label{eq:rimom_vex_zo}
Z_\mathcal{O}^{\MOM}=Z_V^\MOM\mathop{\text{lim}}\limits_{m_R\rightarrow0}\frac{\text{Tr}[\Lambda_{V,B}(p,p)\gamma_\mu]}{\text{Tr}[\Lambda_{\mathcal{O},B}(p,p)\Lambda^{\text{tree}}_{\mathcal{O}}(p,p)^{-1}]}\Big |_{p^2=\mu^2}.
\end{equation}

In the $\SMOM$ scheme, 
the momenta of external quark legs are systematically set to be
\begin{eqnarray}
p_1^2=p_2^2=(p_2-p_1)^2=\mu^2.
\end{eqnarray}
The renormalization conditions for the quark self energy, scalar, pseudoscalar, tensor, vector and axial vector currents are chosen to be~\cite{Aoki:2007xm,Sturm:2009kb}
\begin{subequations}
\label{eq:rismom}
\begin{eqnarray}
\label{eq:rismom_Zq}
Z_q^{\SMOM}&=&\mathop{\textrm{lim}}\limits_{m_R\rightarrow0}\frac{-i}{12p^2}\textrm{Tr}\Big[S^{-1}_B(p)\slashed{p}\Big]_{p^2=\mu^2},  \\
\label{eq:rismom_ZS}
Z_{S/P/T}^\SMOM&=&\mathop{\text{lim}}\limits_{m_R\rightarrow0}\frac{Z_q^\SMOM}{\frac{1}{12}\text{Tr}[\Lambda_{S/P/T,B}(p_1,p_2)\Lambda^{\text{tree}}_{S/P/T}(p_1,p_2)^{-1}]},\\
\label{eq:rismom_ZV}
Z_V^\SMOM&=&\mathop{\text{lim}}\limits_{m_R\rightarrow0}\frac{Z_q^\SMOM}{\frac{1}{12q^2}\text{Tr}[q_\mu\Lambda^\mu_{V,B}(p_1,p_2)\slashed{q}]},\\
\label{eq:rismom_ZA}
Z_A^\SMOM&=&\mathop{\text{lim}}\limits_{m_R\rightarrow0}\frac{Z_q^\SMOM}{\frac{1}{12q^2}\text{Tr}[q_\mu\Lambda^\mu_{A,B}(p_1,p_2)\gamma_5\slashed{q}]}.
\end{eqnarray}
\end{subequations} 
If one use Eq.~(\ref{eq:rismom_Zq}) to calculate the $Z_q^{\SMOM}$ in the lattice, the result will suffer large discretization error~\cite{Chang:2021vvx}. Thus we can follow the similar strategy used by the $\MOM$ scheme, and use the following vector current correction to obtain the $Z_q^{\SMOM}$,
\begin{equation}
\label{eq:smom_vex_zq}
Z_q^{\SMOM}=\mathop{\text{lim}}\limits_{m_R\rightarrow0}\frac{Z_V^\SMOM}{12}\text{Tr}[q_\mu\Lambda^\mu_{V,B}(p_1,p_2)\slashed{q}]_{sym},
\end{equation}
then define RCs of other quark operator as
\begin{equation}
\label{eq:smom_vex_zo}
Z_\mathcal{O}^{\SMOM}=Z_V^\SMOM\mathop{\text{lim}}\limits_{m_R\rightarrow0}\frac{\text{Tr}[q_\mu\Lambda^\mu_{V,B}(p_1,p_2)\slashed{q}]}{\text{Tr}[\Lambda_{\mathcal{O},B}(p_1,p_2)\Lambda^{\text{tree}}_{\mathcal{O}}(p_1,p_2)^{-1}]}\Big |_{p^2=\mu^2}.
\end{equation}

In this work, we use both the $\MOM$ and $\SMOM$ schemes as the intermediate schemes to renormalize the current quark operators on the several gauge ensembles. The gauge ensembles use the dynamical Domain wall fermion~(DWF) action and Symanzik gauge action~\cite{RBC:2014ntl,Boyle:2015exm} with the bare guage coupling $6/g^2\sim 2$, or 
Highly improved stagger quark~(HISQ) actions and Iwasaki gauge action ~\cite{MILC:2012znn} with the bare guage coupling $6/g^2\sim 4$.
The information of these gauge ensembles are listed in the Table.~(\ref{tab:conf}).

To suppress the discretization error, the momenta are chosen to be close the body diagonal in the $\MOM$ scheme, i.e.,
\begin{eqnarray}\label{eq:demo_cond}
\frac{p^{[4]}}{(p^2)^2}<0.28,  \text{where}~ p^{[4]}=\sum\limits_\mu p^4_\mu,~p^2=\sum\limits_{\mu}p^2_\mu.
\end{eqnarray}
In the $\SMOM$ scheme, the momenta is chosen to be symmetrical form, 
such as $p_1$=($q$,$q$,0,0), $p_2$=(0,$q$,$q$,0).
The results in the $\MOM$ and $\SMOM$ can be converted to the $\MSbar$ scheme by using the perturbative matching factors, which 
can be found in ~\cite{Franco:1998bm, Chetyrkin:1999pq, Almeida:2010ns,Gracey:2011fb,Kniehl:2020sgo}. 
Then we use the anomalous dimensions to evolve the results to 2 GeV. The final results can be obtained by applying
appropriate ansatz to extrapolate the results to $a^2p^2\rightarrow0$ limit. In Fig.~\ref{fig:ZS_msbar}, 
we present our results about $Z_S^\MSbar(2\GeV)$ on the different 
gauge ensembles from $\MOM$ (red data points) and $\SMOM$ schemes (blue data points).
The rectangle and circle represent the results on the HISQ ensemble and DWF ensembles. We can see that the result is a smooth function of the lattice spacing, rather than $6/g^2$ which can be sensitive to the used gauge action. 

\makeatletter\def\@captype{table}\makeatother
\begin{minipage}{.4\textwidth}
\centering
\setlength{\tabcolsep}{0.5mm}{
  \begin{tabular}{l|lcrlc}
\hline
\hline
tag &  $6/g^2$ & $L$ & $T$ & $a(\mathrm{fm})$ & $m_{\pi}$ (MeV) \\
\hline
HISQ12 &  3.60 & 24 & 64 & 0.1213(9) & 310\\
\hline
HISQ09 &  3.78 & 32 & 96 & 0.0882(7)& 310\\
\hline
HISQ06 &  4.03 & 48 & 144 & 0.0574(5)& 310 \\
\hline
HISQ04 &  4.20 & 64 & 192 & 0.0425(4)& 310 \\
\hline
24D &  1.633  & 24 & 64 & 0.194(2) & 139 \\
\hline
24DH & 1.633 & 24 & 64 & 0.194(2) & 337  \\
\hline
32Dfine & 1.75  & 32 & 64 & 0.143(2) & 139  \\
\hline
48I &  2.13 & 48 & 96 & 0.1141(2) &139  \\
\hline
64I &  2.25 & 64 & 128 & 0.0837(2) &139\\
\hline
48If &  2.31 & 48 & 96 & 0.0711(3) &280 \\
\hline
32If &  2.37 & 32 & 64 & 0.0626(4) & 371 \\
\hline
  \end{tabular}}
  \label{tab:conf}
  \caption{Setup of the ensembles, including the bare coupling constant $g$, lattice size $L^3\times T$, 
  lattice spacing $a$ and sea pion mass $m_{\pi}$. }
\end{minipage}
\makeatletter\def\@captype{figure}\makeatother
    \begin{minipage}{0.6\textwidth}
    \setcaptionwidth{3in}
    \centering
    \includegraphics[width=8cm,height=6cm]{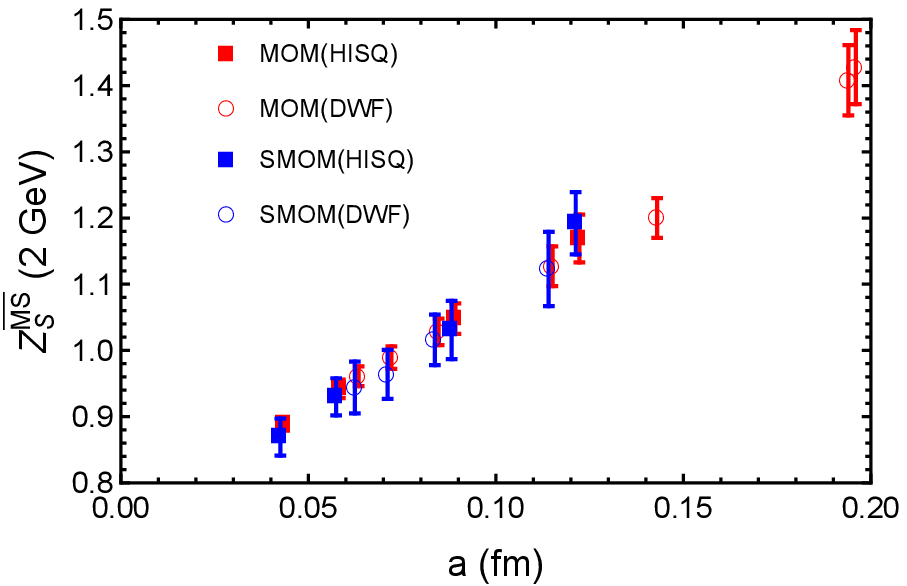}
    \caption{$Z_S^\MSbar(2 \GeV)$ calculated with intermediate $\MOM$ and $\SMOM$ results on the 
    different gauge ensembles we used. }
     \label{fig:ZS_msbar}
    \end{minipage}

\section{Numerical details}

In this section, we choose the RCs on the 48I ensemble to show the updated analysis with a larger $p^2$ range, as those RCs have been calculated in the previous $\chi$QCD work~\cite{Bi:2017ybi}.

\subsection{Renormalization of vector current and axial vector current}
\begin{figure}[!tb]
    \centering
    \includegraphics[height=.22\textheight]{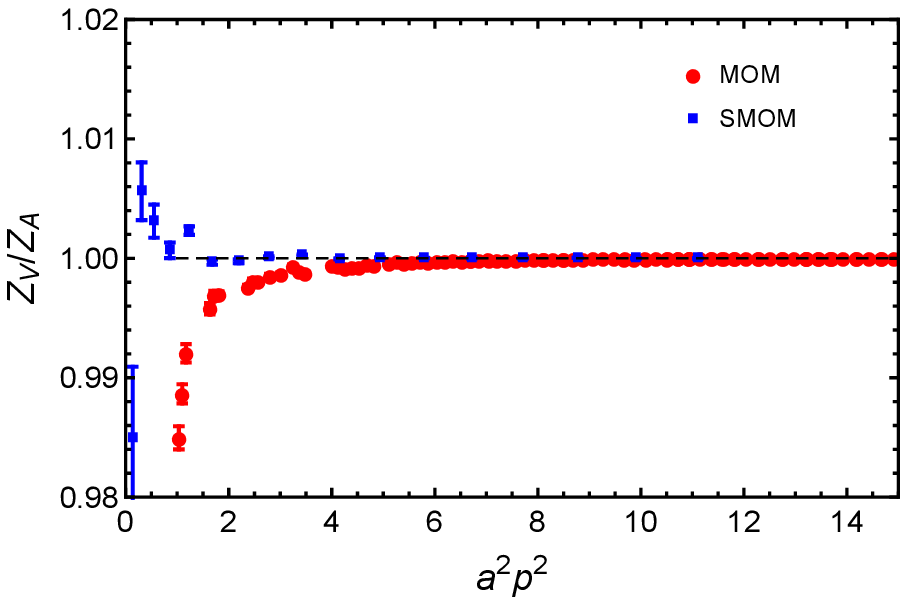}
    \caption{ The ratio $Z_V/Z_A$ in the chiral limit in the $\MOM$ and $\SMOM$ schemes.}
    \label{fig:ZVZA}
\end{figure}

The RC of axial vector current can be calculated through the PCAC relation,
\begin{equation}
Z_A\partial_\mu (\bar{\psi}\gamma_5\gamma_\mu\psi)=2Z_mZ_Pm_q\bar{\psi}\gamma_5\psi=2m_q\bar{\psi}\gamma_5\psi,    
\end{equation}
since $Z_mZ_P$=1 for the overlap fermions. 
In Fig.~\ref{fig:ZVZA}, we present the ratio of $Z_V/Z_A$ in the $\MOM$ and $\SMOM$ schemes,
and it is clear that $Z_V/Z_A=1$ are well satisfied both in the $\MOM$ and $\SMOM$ schemes, as we expect for the chiral fermion. Except the statistical error,
we also estimate systematic errors cause by the finite volume 
effect and non-zero strange quark in the gauge ensemble, more details can be found in our upcoming paper~\cite{He:RC}. The final results of $Z_A$ on different ensembles are shown in Table.~\ref{tab:ZA}.

\begin{table*}[htbp]
  \centering
  \begin{tabular}{c|ccccccccc}
Ensemble & HISQ12 & HISQ09 & HISQ06 & HISQ04 \\
\hline
$Z_A$ & 1.1102(2)(1)(18) & 1.0834(1)(1)(20) & 1.0617(1)(1)(20)  & 1.0523(1)(1)(19) \\
\hline
Ensemble &  48I & 64I & 48If & 32If  \\
\hline
$Z_A$ & 1.1037(1)(1)(21) & 1.0787(1)(1)(19) &  1.0700(1)(1)(18) & 1.0646(2)(2)(19)   \\
\hline
Ensemble &  24D & 24DH & 32Dfine &   \\
\hline
$Z_A$ & 1.2193(3)(1)(27) & 1.2251(3)(1)(27) &  1.1417(2)(1)(20) & 
\\ 
\hline
  \end{tabular}
  \caption{The renormalization constants of the axial current on the different ensembles. The values in the three brackets following the center value correspond to the statistical error, systematic errors caused by the finite volume effect and non-zero strange quark mass.}
  \label{tab:ZA}
\end{table*}

%
\subsection{Renormalization of quark self energy }\label{sec:zq}
\begin{figure}[]
\begin{center}
\subfigure[]
{
	\begin{minipage}[b]{0.45\linewidth}
	\centering 
	\includegraphics[scale=0.7]{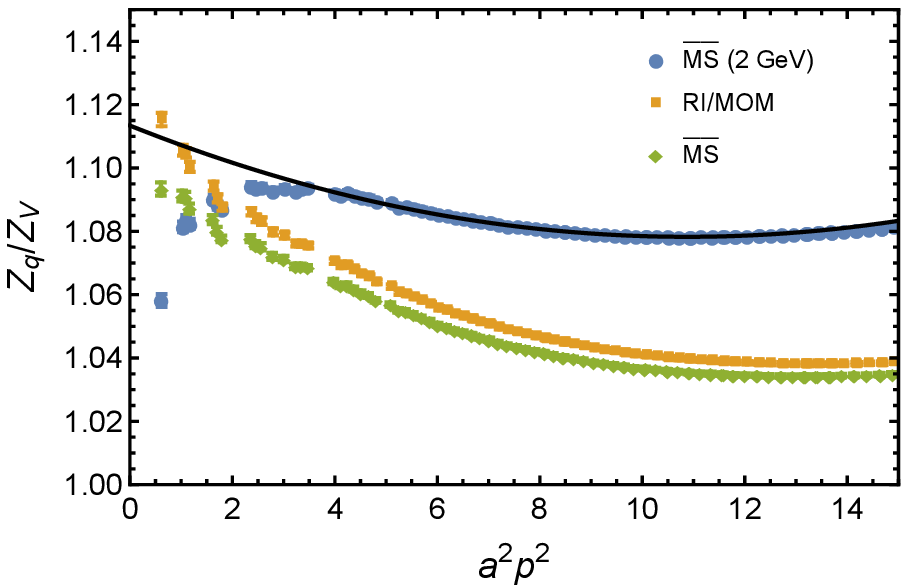}
	\end{minipage}
}
\subfigure[]
{
	\begin{minipage}[b]{0.45\linewidth}
	\centering  
	\includegraphics[scale=0.7]{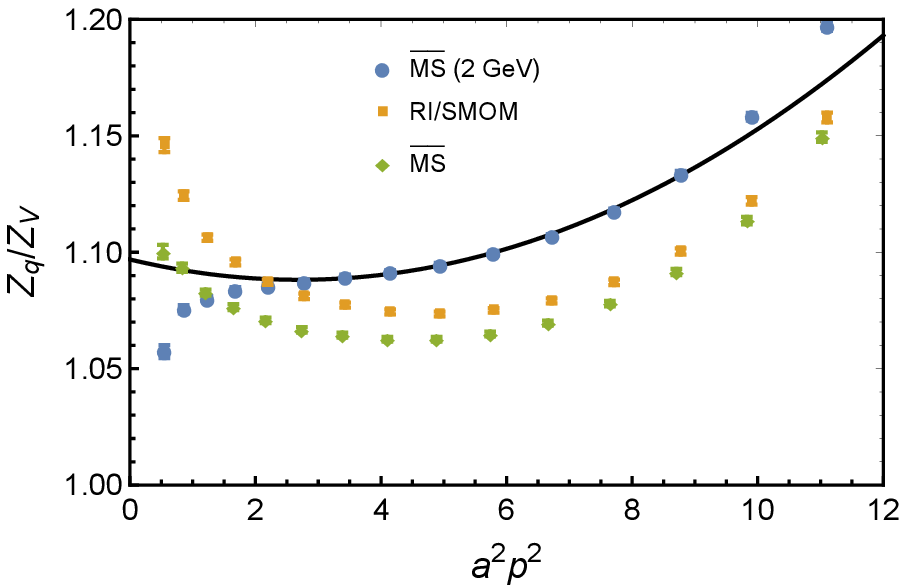}
	\end{minipage}
}
\caption{The conversion of $Z_q/Z_V$ in the $\MOM$ scheme (left panel) and $\SMOM$ scheme (right panel) to the $\MSbar$ schemes.
The results at 2GeV are obtained using the anomalous dimension of quark field strength in the $\MSbar$ scheme. 
The solid lines represent the extrapolation of $a^2p^2$ with Eq.~(\ref{eq:fit_model1}) using the data in the specific region
we mention in the text. }
\label{fig:Zq_result}
\end{center}
\end{figure}

The RCs of quark field strength in the $\MOM$ and $\SMOM$ schemes can be 
obtained the vector current vertex correction in Eq.~(\ref{eq:rimom_vex_zq}) and Eq.~(\ref{eq:smom_vex_zq}). 
Using perturbative matching factors, one can convert the RCs in the intermediate schemes 
to the $\MSbar$ scheme. These results are presented in Fig.~(\ref{fig:Zq_result}).  To remove the discretiztion error,
we use the following ansatz to fit the data
\begin{equation}\label{eq:fit_model1}
f=c_0+c_1a^2p^2+c_2(a^2p^2)^2,
\end{equation}
the fit regions for the results from $\MOM$ scheme and $\SMOM$ are chosen to be $a^2p^2\in[6:12]$ and $a^2p^2\in[2.5:9]$, respectively.
The corrsepoding fit results for $c_0$ are 1.1142(27) and 1.0969(38).  
In addition to the statistical error, 
we should also consider the systematical error caused by the fit region of $a^2p^2$, 
perturbative matching factors, finite volume effect, non-zero strange quark 
and the uncertainties of $\Lambda_{QCD}$ and lattice spacing, etc,
and the details will be presented in~\cite{He:RC,Bi:2017ybi}. 
The final results for the $Z_q^\MSbar(2~\GeV)/Z_V$ from the intermediate $\MOM$ and $\SMOM$ are consistent, which are 1.114(5) and 1.097(23).
The most uncertainties in the $\MOM$ and $\SMOM$ schemes are from the truncation error in the perturbative 
matching from the $\MOM$ scheme to $\MSbar$ scheme and the fit region of $a^2p^2$ in the $a^2p^2$ extrapolation, respectively.

\subsection{Renormalization of scalar and pseudoscalar quark operator}
\begin{figure}[]
\begin{center}
\subfigure[]
{
	\begin{minipage}[b]{0.45\linewidth}
	\centering 
	\includegraphics[scale=0.7]{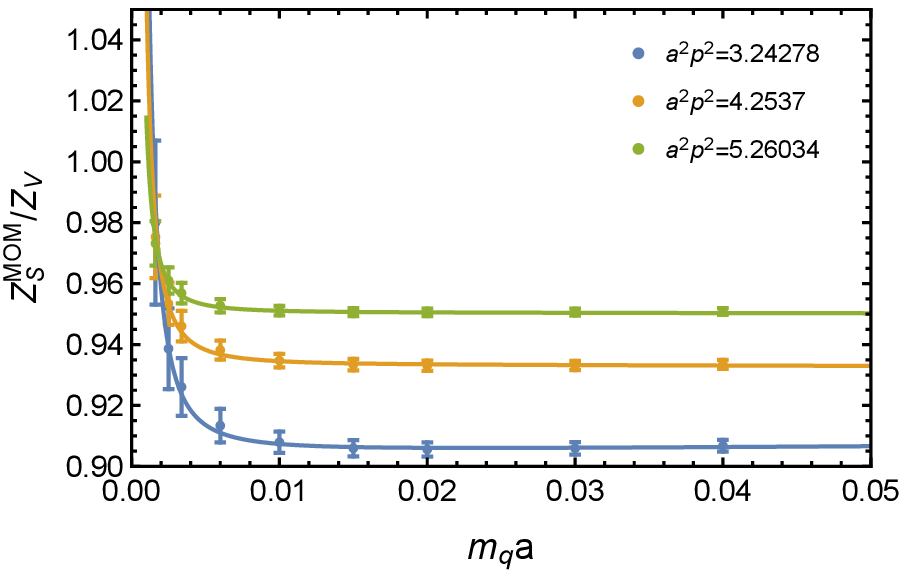}
	\end{minipage}
}
\subfigure[]
{
	\begin{minipage}[b]{0.45\linewidth}
	\centering  
	\includegraphics[scale=0.7]{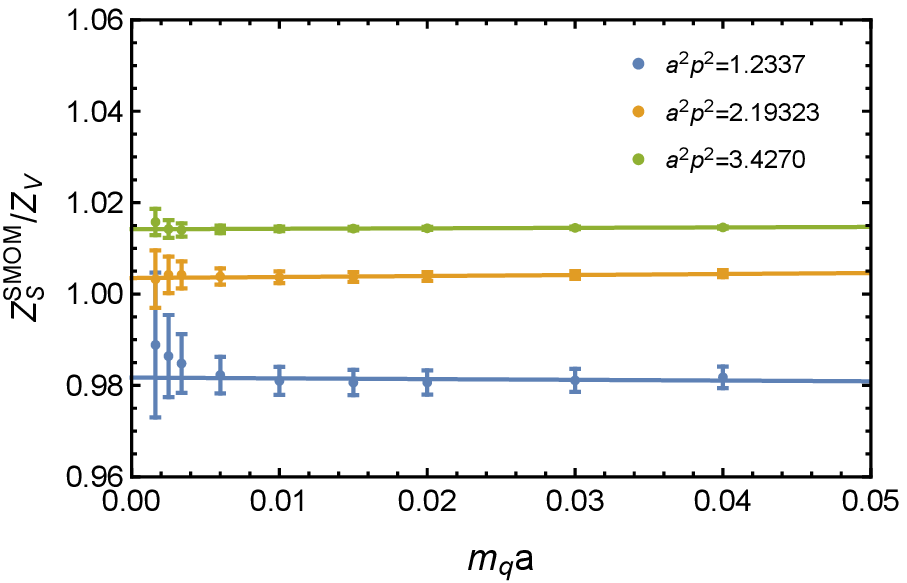}
	\end{minipage}
}
\subfigure[]
{
	\begin{minipage}[b]{0.45\linewidth}
	\centering    
	\includegraphics[scale=0.7]{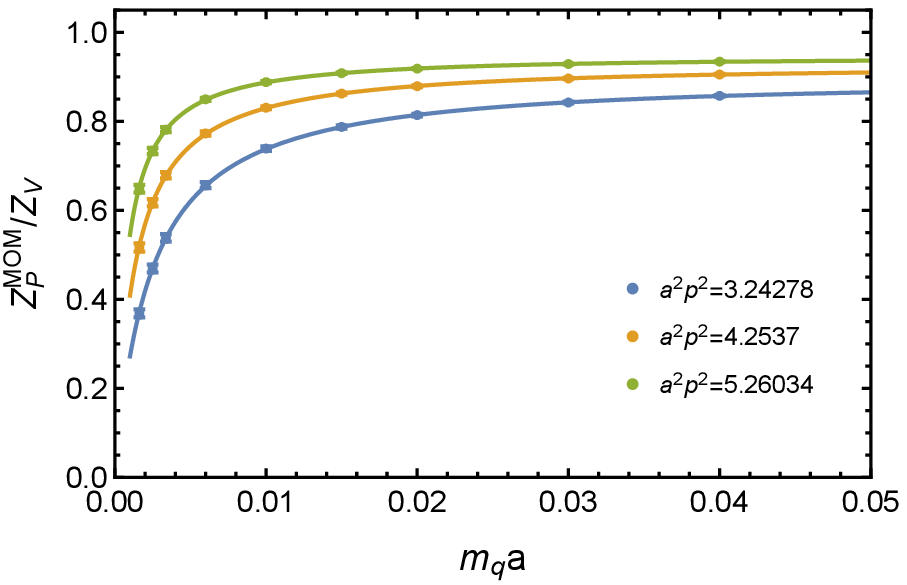} 
	\end{minipage}
}
\subfigure[] 
{
	\begin{minipage}[b]{0.45\linewidth}
	\centering      
	\includegraphics[scale=0.7]{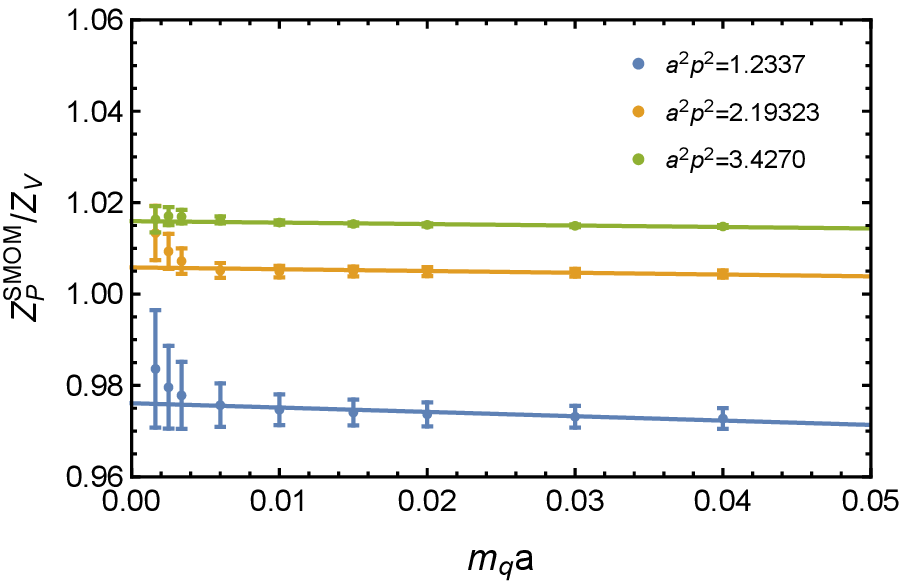} 
	\end{minipage}
}
\caption{The quark mass dependence for the $Z_S/Z_V$ and $Z_{P}/Z_V$ in the $\MOM$ scheme and 
$\SMOM$ schemes.}
\label{fig:ZsZpmass}
\end{center}
\end{figure}

There are unphysical mass poles in the forward scalar and pseudoscalar quark matrix elements,
they correspond to the contribution from the zero mode of Dirac operator~\cite{Blum:2001sr} and 
goldstone mass pole~\cite{Martinelli:1994ty} in the chiral limit.
In Fig.~\ref{fig:ZsZpmass}, we present the quark mass dependence for the RCs of scalar quark and
pseudoscalar quark operators in the $\MOM$ and $\SMOM$ schemes. One can see that
$Z_S^\MOM$ and $Z_P^\MOM$ are very sensitive to the valence quark mass.
We use the following ansatz to extrapolate the results to the chiral limit,
\begin{equation}
Z_S^\MOM/Z_A(am_q)=\frac{A_s}{(am_q)^2}+B_s+C_sam_q, ~~~~~~~Z_P^\MOM/Z_A(am_q)=1/\Big\{\frac{A_p}{(am_q)}+B_p+C_pam_q\Big\},
\end{equation}
where $B_s$ and $1/B_p$ are the results of $Z_S^\MOM/Z_V$ and $Z_P^\MOM/Z_V$ in the chiral limit.
The contamination of both the zero mode and goldstone pole are much smaller in the $\SMOM$ scheme, it allows us to choose the linear chiral extrapolation. 

\begin{figure}[]
\begin{center}
\subfigure[]
{
	\begin{minipage}[b]{0.45\linewidth}
	\centering 
	\includegraphics[scale=0.7]{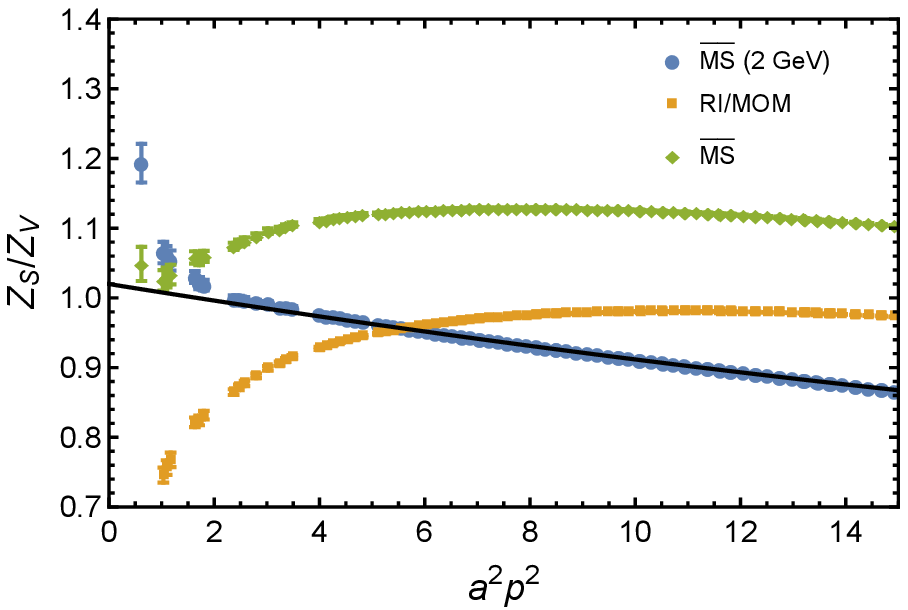}
	\end{minipage}
}
\subfigure[]
{
	\begin{minipage}[b]{0.45\linewidth}
	\centering  
	\includegraphics[scale=0.7]{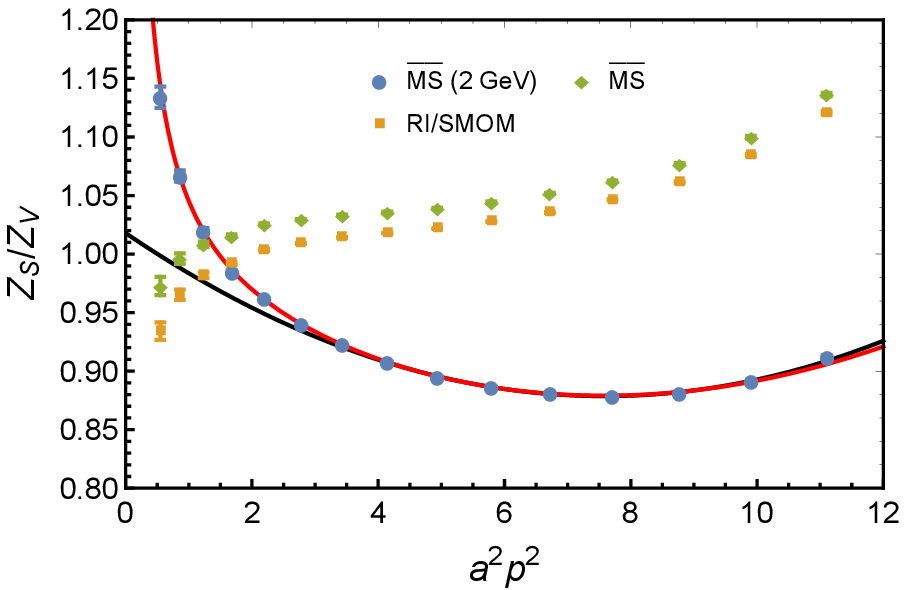}
	\end{minipage}
}
\caption{Conversion and running of $Z_S/Z_V$ for the intermediate schemes $\MOM$ and $\SMOM$ schemes.
The black curves are the fittings using Eq.~(\ref{eq:fit_model1}), the red curve in the right panel is the fitting with Eq.~(\ref{eq:fit_model2}).}
\label{fig:Zs_result}
\end{center}
\end{figure}

The conversion of $Z_S/Z_V$ from the intermediate schemes to the $\MSbar$ scheme are presented in Fig.~\ref{fig:Zs_result}.  
One can see that the $\SMOM$ scheme has better convergence of the perturbative matching than the $\MOM$ scheme.
However, after converting to the $\MSbar$ and running to 2 GeV, the results from the $\SMOM$ scheme have 
very non-linear dependence on $a^2p^2$, but
that $Z_S^\MSbar(2\GeV)/Z_A$ calculated 
with the $\MOM$ scheme are almost linear on $a^2p^2$ when $4\leq a^2p^2$.   
Using the fit ansatz in Eq.~(\ref{eq:fit_model1}) to fit the data in $a^2p^2\in[6:12]$ from the $\MOM$ scheme, we obtain $Z_S^\MSbar(2 \GeV)/Z_A$ is 1.0208(29).
When we use Eq.~(\ref{eq:fit_model1}) to fit the results from the $\SMOM$ scheme, we find
the largest fit region we can use is $a^2p^2\in[4:9]$ if the upper limit is fixed and the constraint $\chi^2/d.o.f\le 1$ is applied. 
The corresponding fit result is $Z_S^\MSbar(2\GeV)/Z_A=1.0179(25)$.  
In addition, we also consider the following empirical form to fit the result from the $\SMOM$ scheme,
\begin{equation}\label{eq:fit_model2}
f=\frac{c_1}{a^2p^2}+c_0+c_1a^2p^2+c_2(a^2p^2)^2,
\end{equation}
then the fit range can be extended to be $a^2p^2\in[1:9]$ with $\chi^2/d.o.f=0.6$ after $1/(a^2p^2)$ term is included,
and the corresponding fit result for $c_0$ is 0.9652(44). We take
the deviation between results obtained by two different fit models as the systematic error.  We also considered the uncertainty 
caused by the other sources as we mention in the Sec.~(\ref{sec:zq}). The final results for $Z_S^\MSbar$ is 1.127(30) and 1.123(58) 
from the $\MOM$ and the $\SMOM$ schemes. The most uncertainty is contributed by the 
truncation error in the matching factor for the $\MOM$ scheme and 
the deviation between different fit models for the $\SMOM$ scheme.

\subsection{Renormalization of tensor quark operator }
\begin{figure}[]
\begin{center}
\subfigure[]
{
	\begin{minipage}[b]{0.45\linewidth}
	\centering 
	\includegraphics[scale=0.75]{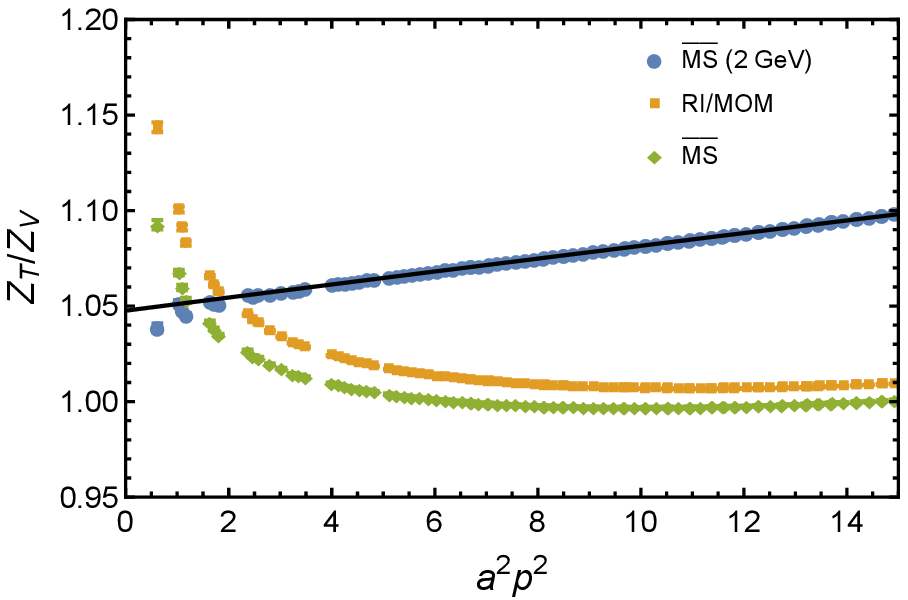}
	\end{minipage}
}
\subfigure[]
{
	\begin{minipage}[b]{0.45\linewidth}
	\centering  
	\includegraphics[scale=0.75]{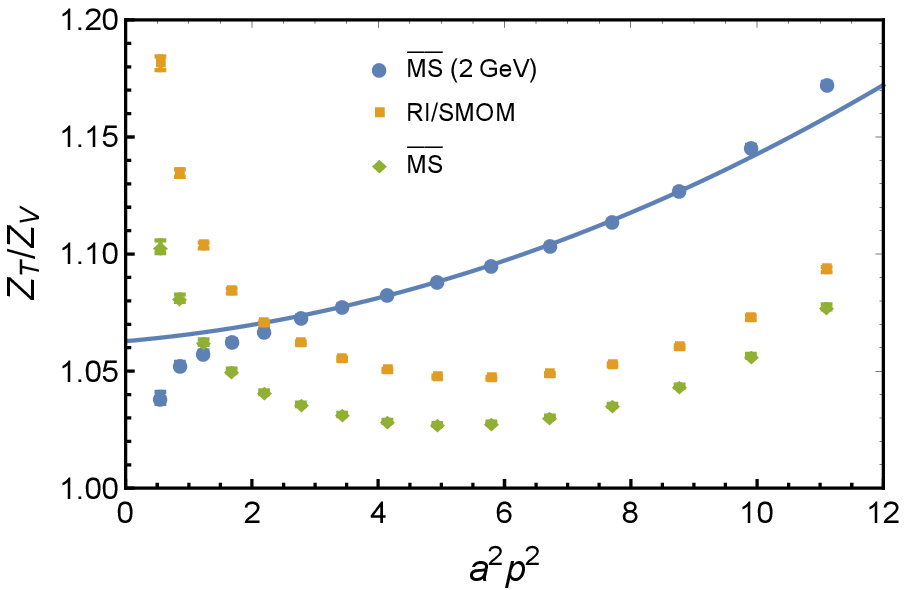}
	\end{minipage}
}
\caption{Similar as Fig.~\ref{fig:Zq_result} but for $Z_T/Z_V$.
}
\label{fig:Zt_result}
\end{center}
\end{figure}

The results of $Z_T^\MSbar(2 \GeV)/Z_V$ from the $\MOM$ and $\SMOM$ are presented in Fig.~\ref{fig:Zt_result}.
The results from $\MOM$ scheme show a linear dependence on the $a^2p^2$. And we also choose Eq.~(\ref{eq:fit_model1}) to
fit data in $a^2p^2\in[6:12]$ and $a^2p^2\in[2.5:9]$ for the results from the $\MOM$ and $\SMOM$ schemes, respectively. The fit results we obtained are
1.0486(5) and 1.0628(23). Contracted to the scalar operator case, $Z_T$ using $\MOM$ scheme shows a better perturbative convergence than that using the $\SMOM$ scheme. After considering all the systematic errors, 
our results about $Z_T^\MSbar(2 \GeV)$ is about 1.158(3) and 1.173(18). The uncertainty of results from $\MOM$ scheme is small
and the most uncertainty in the result from $\SMOM$ scheme is caused by the different fit region of $a^2p^2$.

\section{Conclusion}
In this work, we renormalize the bare matrix elements of vector, axial vector, scalar, pseudoscalar and tensor quark operators calculated with 
valence overlap fermion on the dynamical DWF and HISQ gauge ensembles. Using the perturbative matching factors and 
anomalous dimensions, we convert the results from the intermediate $\MOM$ and $\SMOM$ schemes
to the $\MSbar$ 2 $\GeV$. In addition the statistical error, we also 
consider kinds of systematic errors in the final results. We find the intermediate $\MOM$ and $\SMOM$
schemes can provide consistent results after converting to the $\MSbar$ scheme and doing appropriate $a^2p^2$ extrapolation.
 The $\SMOM$ scheme show better perturbative convergence when 
matching to the $\MSbar$ scheme for the scalar quark operator,  the case is opposite for the tensor quark operator, 
the perturbative series have better convergence for the $\MOM$ scheme.  After converting the results to 2 GeV, the 
results from the $\MOM$ scheme show better linear dependence on $a^2p^2$. The results from the $\SMOM$ scheme show 
obvious non-linear dependence on $a^2p^2$, it leads to a huge dependence on the  $a^2p^2$ fit region. For the scalar operator,
we use two different fit models to describe the data and fit results between these fit models have obvious deviation, while the deviation decreases slowly on the lattice spacing.

In Table~\ref{tab:results_sum}, we list the RCs of different quark operators on the 11 gauge ensembles.  We would like to mention that the lattice spacing of 24D, 24DH and 32Dfine 
are too large to calculate in $\SMOM$ scheme. 
 One can see that the final results from the $\MOM$ and $\SMOM$ schemes are all consistent
within the uncertainties. 

\section{Ackonwledge}
We thank the RBC and UKQCD collaborations for providing us their DWF gauge configurations, and MILC collaboration for providing the HISQ gauge configurations. The calculations were performed using the GWU code~\cite{Alexandru:2011ee,Alexandru:2011sc} through HIP programming model~\cite{Bi:2020wpt}.  The numerical calculation is supported by the Strategic Priority Research Program of Chinese Academy of Sciences, Grant No. XDC01040100, and also the supercomputing system in the Southern Nuclear Science Computing Center (SNSC). 

\begin{table}[htbp]
  \centering
  \setlength{\tabcolsep}{1mm}{
  \begin{tabular}{l|lccc|cccc}
\hline
\hline
& \multicolumn{4}{c}{Results from the $\MOM$ scheme}   &\multicolumn{4}{|c}{Results from the $\SMOM$ scheme} \\
\hline
Ensemble & $Z_q(2\GeV)$ &$Z_S(2\GeV)$ & $Z_P(2\GeV)$ & $Z_T(2\GeV)$  & $Z_q(2\GeV)$ &$Z_S(2\GeV)$ & $Z_P(2\GeV)$ & $Z_T(2\GeV)$ \\
\hline
HISQ12 &1.246(7)  & 1.169(36) &  1.230(63) & 1.159(3)  &1.230(20)  & 1.192(47) &  1.151(41) & 1.173(15)\\
\hline
HISQ09 &1.204(5)  & 1.048(23) &  1.062(28) &  1.152(3) &1.184(28) & 1.031(44) &  1.028(38) & 1.160(22) \\
\hline
HISQ06 &1.169(7)  & 0.942(14) &  0.944(22) & 1.156(3)   &1.163(20) & 0.930(28) &  0.931(29) & 1.165(16)  \\
\hline
HISQ04 &1.154(6)  & 0.889(10) &  0.890(18) & 1.163(4)   &1.144(15)  & 0.869(28) &  0.871(26) & 1.167(15)  \\
\hline
24D    & 1.364(24) & 1.408(53) & 1.427(62) & 1.230(7)    & $\overline{~~~}$  & $\overline{~~~}$ & $\overline{~~~}$ & $\overline{~~~}$ \\
\hline
24DH    & 1.369(23) & 1.428(56) & 1.454(75) & 1.236(7)    &$\overline{~~~}$  & $\overline{~~~}$ & $\overline{~~~}$ & $\overline{~~~}$ \\
\hline
32Dfine    & 1.254(13) & 1.200(30) & 1.208(37) & 1.181(2)   & $\overline{~~~}$  & $\overline{~~~}$ & $\overline{~~~}$ & $\overline{~~~}$ \\
\hline
48I    & 1.230(6) & 1.127(30) & 1.140(35) & 1.158(3)  &1.211(25)  & 1.123(58) &  1.127(51) & 1.173(18)  \\
\hline
64I    & 1.198(5) & 1.028(20) & 1.032(26) & 1.152(2)  &1.179(29) & 1.016(38) & 1.017(37) & 1.165(20) \\
\hline
48If   & 1.182(6) & 0.989(17) & 0.997(23) & 1.152(2) & 1.168(23) &0.964(37) &0.966(41)  & 1.163(18) \\
\hline
32If   & 1.169(6) & 0.961(15) & 0.964(21) & 1.155(3)  &1.156(15) &0.944(39) & 0.944(44) &1.160(12)  \\
\hline
  \end{tabular}}
  \caption{The RCs of quark sele energy, scalar operator, pesudoscalar operator and tensor operator in $\MSbar$ scheme and 2 $\GeV$ from the $\MOM$ and 
  $\SMOM$ schemes.}
  \label{tab:results_sum}
\end{table}

\providecommand{\href}[2]{#2}\begingroup\raggedright\endgroup

\end{document}